\documentclass[twocolumn,amsmath,amssymb,pra,superscriptaddress]{revtex4-1}
\usepackage[dvipdfmx]{graphicx}
\usepackage{dcolumn}
\usepackage{bm}
\usepackage{subfigure}
\usepackage{booktabs}
\usepackage{color}
\newcounter{one}
\def\ket#1{\mbox{\boldmath $#1$}}
\newcommand{\bracket}[1]{\left\langle #1 \right\rangle}

\newcommand{\affA}{
Artificial Intelligence Research Center, 
National Institute of Advanced Industrial Science and Technology, 
2-3-26 Aomi, Koto-ku, Tokyo, Japan
}
\newcommand{\affB}{
Department of Mathematical and Computing Science, Tokyo Institute of Technology, 
W8-45, 2-12-1 Ookayma, Meguro-ku, Tokyo, Japan
}

\begin{document}

\title{\textbf{Counting the number of metastable states in the modularity landscape: Algorithmic detectability limit of greedy algorithms in community detection}}

\author{Tatsuro Kawamoto}
\affiliation{\affA}
\author{Yoshiyuki Kabashima}
\affiliation{\affB}
\date{\today}

\begin{abstract}
Modularity maximization using greedy algorithms continues to be a popular approach toward community detection in graphs, even after various better forming algorithms have been proposed. 
Apart from its clear mechanism and ease of implementation, this approach is persistently popular because, presumably, its risk of algorithmic failure is not well understood. 
This Rapid Communication provides insight into this issue by estimating the algorithmic performance limit of modularity maximization. 
This is achieved by counting the number of metastable states under a local update rule. 
Our results offer a quantitative insight into the level of sparsity at which a greedy algorithm typically fails. 
\end{abstract}
 
\maketitle

\textit{Introduction}--- 
Since the proposal of the modularity function \cite{NewmanGirvan2004}, a number of its maximization algorithms and related objective functions have been put forward, and some of them have been widely applied to the discovery of community structures in real-world networks \cite{Fortunato2010}. 
Modularity maximization is also known to be equivalent to the maximum likelihood method of a statistical model \cite{Newman2013,Newman2016Equivalence}. 
The corresponding greedy algorithms, such as the \textit{Louvain algorithm} \cite{Blondel2008}, are commonly used for optimization. 
However, greedy algorithms have often been employed as baselines in benchmark tests and various better performing algorithms have been proposed. 
Moreover, from a Bayesian viewpoint \cite{Nishimori2001,Iba1999}, modularity maximization is known to be suboptimal when a graph is generated from an assumed statistical model, which implies the risk of overfitting \cite{ZhangMoore2014}. 
Nevertheless, greedy algorithms remain very popular partly because, 
presumably, we do not know in which cases we should not expect greedy algorithms to work. 

We conducted a theoretical performance analysis to provide insight into this issue. 
In this Rapid Communication, we considered a random graph model with a planted modular structure, called the \textit{stochastic block model} \cite{holland1983stochastic,WangWong87,KarrerNewman2011}, 
which is a canonical model for a theoretical investigation with regard to community detection. 
We derive the limit of the model parameters beyond the point at which a greedy algorithm completely loses the ability to identify the planted modular structure.  
Such a limit is termed as the \textit{algorithmic detectability limit} \cite{DetectabilityTerminology}.
The corresponding limits of other algorithms, e.g., spectral clusterings \cite{NadakuditiNewman2012,Zhang2014,Radicchi2013,KawamotoKabashimaPRE2015,KawamotoKabashimaEPL2015}, and the expectation-maximization algorithm \cite{ADT-Lett,ADT-Full} have also been investigated. 
In contrast, the limit where all algorithms fail is known as the information-theoretic limit \cite{Decelle2011,Decelle2011a,Mossel2015,Massoulie2014}. 
Such a limit exists because, when a planted modular structure is too weak, the corresponding graph instances can also be typically generated by a uniform random graph model. 

Note that, in this Rapid Communication, the structure specified by the planted group assignments is the only community structure defined. 
Although we consider algorithms that aim to maximize modularity, we do not regard the group assignments that achieve the true maximum as the ``real'' community structure. (See the Supplemental Material for further discussions \cite{SupplementalMaterial}.)

Benchmark tests are an experimental approach toward investigating the detectability limits \cite{LFRbenchmark2009,Ronhovde2012,Hu2012,Darst02102014}. 
Although such tests have the advantage of being conducted in a straightforward manner, a definite conclusion can rarely be obtained. 
For example, it is usually unclear whether a better implementation of the adopted algorithm can significantly improve performance, or if there is no hope of improvement because the difficulty is inherent in the formulation. 
By contrast, although theoretical investigations \cite{MooreReview2017,AbbeReview2017} with regard to the detectability limit are available only for limited situations, they enable a more concrete understanding regarding the feasibilities and limitations of community detection. 

In this study, we considered sparse undirected graphs without self-loops or multi-edges. 
With regard to a planted modular structure, we focused on the community structure, i.e., the assortative structures of two groups. 
We define a graph as $G = (V,E)$, where $V$ and $E$ are the sets of vertices and edges, respectively. 
We let $N = |V|$ and the average degree be $c$. 
We denote $\sigma \in \{1, 2\}$ as a group label and $\sigma_{i}$ as the group assignment of vertex $i$. 
We also denote the set of vertices in group $\sigma$ as $V_{\sigma}$, i.e., $\cup_{\sigma} V_{\sigma} = V$ and $\gamma_{\sigma} \equiv |V_{\sigma}|/N$.

\begin{figure}[t!]
 \begin{center}
   \includegraphics[width= \columnwidth]{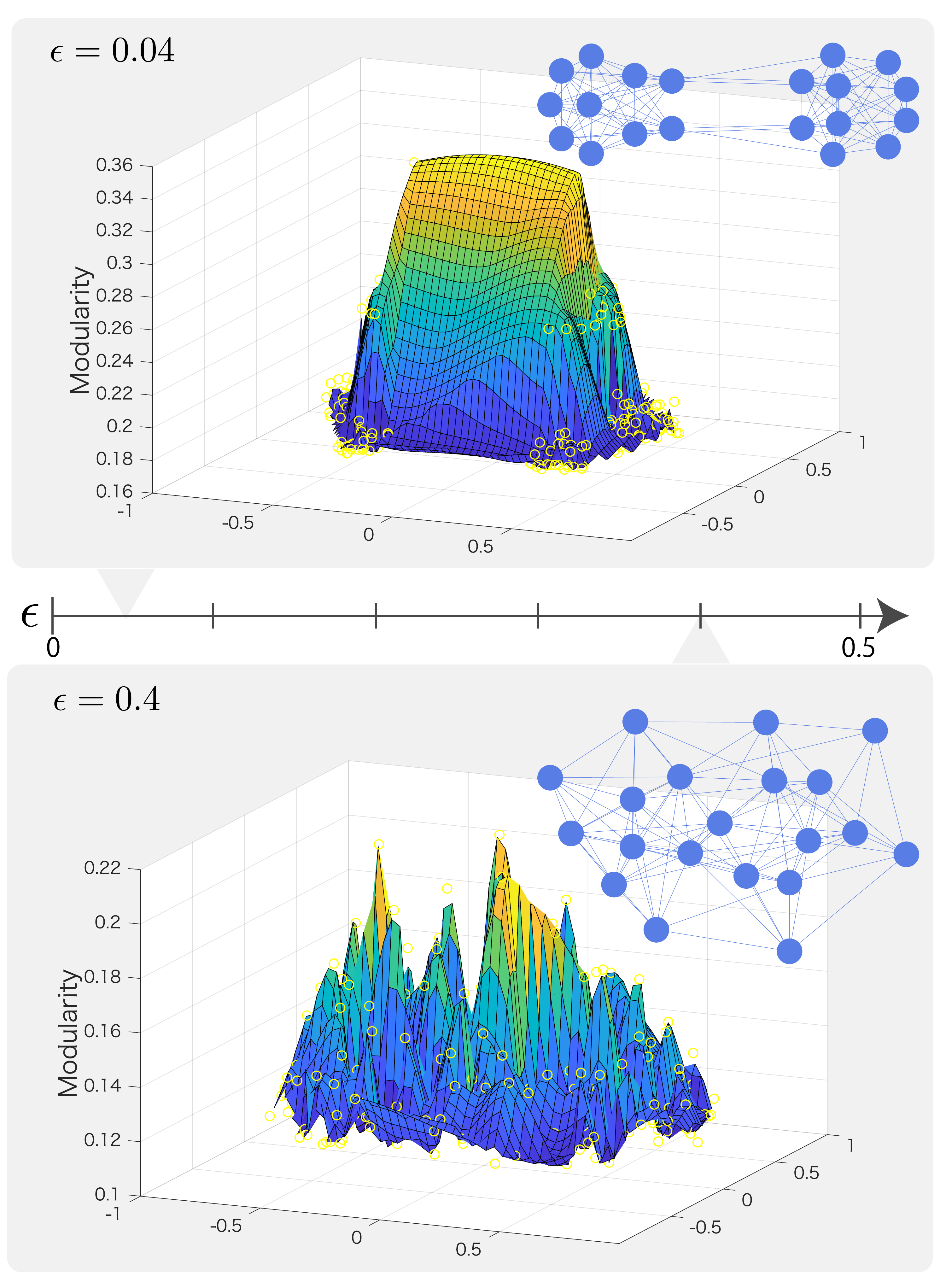}
 \end{center}
 \caption{
	(Color online) 
	Modularity landscapes and network figures for instances of the small stochastic block model with equal group sizes ($N = 20$). 
	The average degrees and strengths of the modular structures are $(c = 9.6, \epsilon = 0.04)$ (top) and $(c = 9.2, \epsilon = 0.4)$ (bottom), respectively. 
	The landscapes were drawn using the code of the curvilinear component analysis distributed by \cite{ClausetCodeWebSite} (see \cite{Good2010} for a detailed description of the visualization).
	}
 \label{SBMlandscape}
\end{figure}

\textit{Stochastic block model}--- 
We denote the adjacency matrix of a graph as $\ket{A}$, where $A_{ij}=1$ when vertices $i$ and $j$ are connected, and $A_{ij}=0$ otherwise. 
The stochastic block model defines the considered probability distribution of the graph configurations, i.e., the graph ensemble. 
In this model, the vertices of a graph have planted group assignments, and the edges are generated independently and randomly on the basis of these assignments. 
For example, the connection probability of the pair of vertices $i$ and $j$ being connected is given by $P_{ij}(A_{ij} = 1) = \rho_{\sigma_{i} \sigma_{j}}$. 
Note that, for the graphs to be sparse, we have $\rho_{\sigma \sigma^{\prime}} = O(N^{-1})$. 
In community detection, $\ket{A}$ is the only input and the objective is to infer the hidden group assignments. 
A particular case wherein the planted group sizes are equal and the connection probability is parametrized as $\rho_{11} = \rho_{22} = \rho_{\mathrm{in}}$ and $\rho_{12} = \rho_{21} = \rho_{\mathrm{out}}$ is often referred to as the symmetric stochastic block model. 
In this case, the strength of the community structure can be parametrized as $\epsilon \equiv \rho_{\mathrm{out}}/\rho_{\mathrm{in}}$.

\textit{Modularity maximization and its detectability}--- 
The objective function of modularity for bipartition can be expressed as
\begin{align}
Q(\ket{s}) = \sum_{ij} s_{i} B_{ij} s_{j} 
= \mathrm{const.} +  \sum_{i,j (i \ne j)} s_{i} B_{ij} s_{j}, \label{Modularity}
\end{align}
where $s_{i} \in \{-1, +1\}$ is a \textit{spin} variable representing the group assignment of vertex $i \in V$, and matrix $\ket{B}$ is defined as 
\begin{align}
B_{ij} \equiv A_{ij} - \alpha c_{i} c_{j}, 
\end{align}
where $c_{i}$ is the degree of vertex $i$ defined as $c_{i} = \sum_{j} A_{ij}$, $\alpha$ is an $O(N^{-1})$ scaling parameter given as the input, and $\alpha$ is called the resolution parameter. 
For a given adjacency matrix $\ket{A}$, the set of most plausible group assignments $\ket{s} = \{s_{1}, \dots, s_{N}\}$ is obtained as that maximizing Eq.(\ref{Modularity}). 

Here, we consider the update of group assignments $\ket{s}$ by a single spin flip, i.e., we may flip only one component $s_{i}$ at each update. 
Note that the global maximum of $Q(\ket{s})$ may not be achieved by a single spin flip, owing to the existence of \textit{metastable states}. 
We define the metastable state as a spin configuration $\ket{s}$ such that $Q(\ket{s})$ does not increase by any single spin flip. 

An intuitive understanding of the algorithmic detectability limit is presented below. 
When Eq.(\ref{Modularity}) does not have metastable states, i.e., local maxima and saddle points, a local update algorithm is able to find its global maximum by starting from an arbitrary random initial state of group assignments. 
Even when metastable states exist, unless their number is sufficiently large, a local update algorithm can still achieve the global maximum of Eq.(\ref{Modularity}) by repeating the algorithm with various initial states. 
However, when the number of metastable states grows exponentially with respect to $N$, it is practically impossible to achieve the global maximum, because a repeated search from an extremely large number of initial states is required. 
Therefore, the detectability limit can be evaluated by counting the number of metastable states. 
To illustrate such a situation, the modularity landscape near the global optimum for the small stochastic block model is shown in Fig.\ref{SBMlandscape} for $\epsilon = 0.04$ and $\epsilon = 0.4$, respectively. As the planted modular structure becomes less clear (larger $\epsilon$), the landscape becomes more ragged. 
In fact, in many real-world networks, modularity landscapes are often very ragged near their global maximum \cite{Good2010}.

\textit{Number of metastable states}--- 
The variation of the objective function $\Delta Q(s_{i})$ caused by the spin flip with respect to $s_{i}$ reads as
\begin{align}
\Delta Q(s_{i}) 
&= (-s_{i}) \sum_{j (\ne i)} B_{ij} s_{j} - s_{i} \sum_{j (\ne i)} B_{ij} s_{j} \notag\\
&= -2 s_{i} \sum_{j (\ne i)} B_{ij} s_{j}. 
\end{align}
Thus, the metastable state is the spin configuration $\ket{s}$ such that $\Delta Q(s_{i}) \le 0$ for all $i$. 
In other words, it is either $s_{i} = \mathrm{sgn}\left( \sum_{j (\ne i)} B_{ij} s_{j} \right)$ or $\sum_{j (\ne i)} B_{ij} s_{j} = 0$. 
This condition can also be expressed such that there exists a non-negative value of $\lambda_{i}$ for each $i \in V$ such that 
\begin{align}
\lambda_{i} = s_{i} \sum_{j (\ne i)} B_{ij} s_{j}. \label{MetastableCondition}
\end{align}

Based on the observation expressed in Eq.(\ref{MetastableCondition}), the number of metastable states $\mathcal{N}_{m}$ can be counted as follows \cite{TanakaEdwards1980}. 
\begin{align}
\mathcal{N}_{m} &= \sum_{\{s_{i}\}} \prod_{i} \int_{0}^{\infty} d\lambda_{i} \, \delta\left( \lambda_{i} - s_{i} \sum_{j (\ne i)} B_{ij} s_{j} \right), \label{Nmetastable1} 
\end{align}
where $\delta(\cdot)$ is Dirac's delta function. 

We are interested in the typical number of metastable states within the graph ensemble, rather than a single graph instance. 
To this end, we estimate the configuration average of graphs generated by the stochastic block model in the limit of $N\to\infty$, which we denote as $\left[ \mathcal{N}_{m} \right]_{A}$. 
However, its exact calculation is technically difficult. 
Therefore, we adopt the \textit{rotating wave approximation} \cite{FujiiRWA2017} or low frequency approximation for the contribution from the delta functions in Eq.(\ref{Nmetastable1}). 
By conducting the calculation described in the Supplemental Material \cite{SupplementalMaterial}, we arrive at the following expression.
\begin{align}
&\left[ \mathcal{N}_{m} \right]_{A} \sim \mathrm{e}^{N f}. 
\end{align}
Here, instead of the result obtained for a general case (found in the Supplemental Material \cite{SupplementalMaterial}), we show a compact expression obtained by 
considering the symmetric stochastic block model and adopting an approximation such that the graph is regular, i.e., the degree is constant for all vertices. 
Additionally, we let the resolution parameter be $\alpha N = 1/c$, which is often employed as the ``standard value''. 
In such a case, the following relationship holds, 
\begin{align}
&f = -\frac{1}{2c} \hat{E}^{2}_{1} - \frac{1}{c} \frac{1+\epsilon}{1-\epsilon} \tilde{G}_{1} \tilde{F}_{1} + \log Z_{1}. \label{Nmetastable-symmetricSBM}
\end{align}
The subscript $1$ indicates that the variables are values for $\sigma=1$. Because of symmetry, the magnitudes of the variables for $\sigma=2$ are equal to those of $\sigma=1$. 
In Eq.(\ref{Nmetastable-symmetricSBM}), $Z_{\sigma}$ is a function of $\hat{E}_{\sigma}$, $\tilde{F}_{\sigma}$, and $\tilde{G}_{\sigma}$, 
\begin{align}
& Z_{\sigma} = \sum_{s} \mathrm{e}^{s\tilde{F}_{\sigma}} \Phi\left( \frac{1}{\sqrt{c}} (\hat{E}_{\sigma} + s \tilde{G}_{\sigma}) \right),
\end{align}
where $\Phi(\cdot)$ is the standard normal cumulative distribution function, 
\begin{align}
\Phi(y) \equiv \int^{y}_{-\infty} \frac{dx}{\sqrt{2\pi}} \mathrm{e}^{-\frac{1}{2} x^{2}}. 
\end{align}
Therefore, the number of metastable states can be evaluated if $\hat{E}_{1}$, $\tilde{F}_{1}$, and $\tilde{G}_{1}$ are determined. 
According to the saddle-point conditions, these variables are evaluated by the following self-consistent equations. 
\begin{align}
& \hat{E}_{1} = \frac{2c}{Z_{1} \sqrt{2\pi c}} \exp\left[ -\frac{1}{2c} (\hat{E}^{2}_{1} + \tilde{G}_{1}^{2})\right]
\cosh ( \tilde{F}_{1} - c^{-1}\hat{E}_{1}\tilde{G}_{1}), \\
& \tilde{F}_{1} = \frac{\mathcal{G}}{Z_{1}\sqrt{2\pi c}} \exp\left[ -\frac{1}{2c} (\hat{E}^{2}_{1} + \tilde{G}_{1}^{2})\right] \sinh(\tilde{F}_{1} - c^{-1}\hat{E}_{1}\tilde{G}_{1}), \\
& \tilde{G}_{1} = \frac{\mathcal{G}}{2 Z_{1}} \sum_{s}\, s \, \mathrm{e}^{s \tilde{F}_{1}} 
\Phi\left( \frac{1}{\sqrt{c}} (\hat{E}_{1} + s\tilde{G}_{1}) \right).
\end{align}
where $\mathcal{G} \equiv 2c (1 - \epsilon)/(1 + \epsilon) = (\rho_{\mathrm{in}} - \rho_{\mathrm{out}})N$.

\begin{figure}[t!]
 \begin{center}
   \includegraphics[width= \columnwidth]{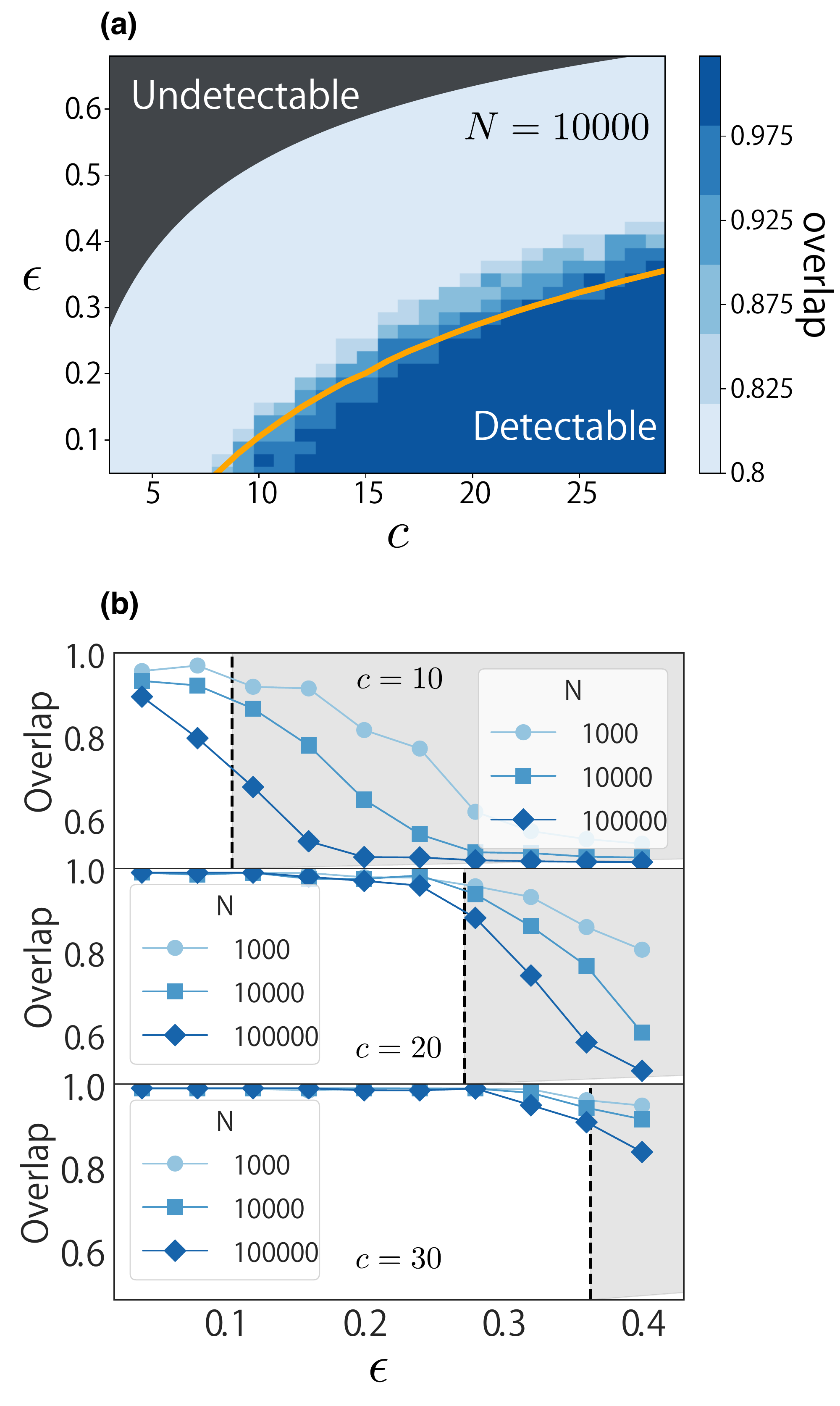}
 \end{center}
 \caption{
	(Color online) 
	(a) Detectability phase diagram of a simple greedy algorithm for the symmetric stochastic block model with $N=10000$. 
	The density plot represents the overlaps, while the solid yellow line represents our detectability limit estimate. 
	The shaded region at the upper-left corner represents the region where the detection is information-theoretically impossible. 
	(b) Overlaps of $c = 10$ (top), $c = 20$ (middle), and $c = 30$ (bottom) as functions of $\epsilon$ for different graph sizes $N$. 
	The shaded region with the dashed border line represents our undetectable region estimate. 
	In all plots, the average overlap value of $100$ graph instances was determined for each pair of $c$ and $\epsilon$ values.
	}
 \label{GreedyPhaseDiagram}
\end{figure}

\textit{Detectability limit of a simple greedy algorithm}--- 
From Eq.(\ref{Nmetastable-symmetricSBM}), it is evident that the graphs have an exponentially large number of metastable states as long as $f > 0$. 
Otherwise, they only have a subexponential number of metastable states. 
Thus, the detectability limit is located at the value of $\epsilon^{\ast}$ where 
\begin{align}
c \log Z_{1} 
= \frac{1}{2} \hat{E}^{2}_{1}  + \frac{1+\epsilon^{\ast}}{1-\epsilon^{\ast}} \tilde{G}_{1} \tilde{F}_{1} \label{TheDetectabilityLimit}
\end{align}
is satisfied. 

The accuracy of our estimate is shown in Fig.\ref{GreedyPhaseDiagram}. 
Here, we consider a simple greedy algorithm, wherein the vertex to be updated is chosen randomly and its spin $s_{i}$ variable is flipped if $\Delta Q(s_{i}) > 0$. 
This algorithm is exactly the process considered in metastable state counting. 
The detectability phase diagram of this algorithm is shown in Fig.\ref{GreedyPhaseDiagram}(a) as a density plot, 
and is obtained by executing the algorithm for the graphs generated by the stochastic block model with various values of the average degree $c$ and strength of community structure $\epsilon$. 
The color depth represents the \textit{overlap}, which is defined as the fraction of vertices correctly assigned to the planted groups, i.e., 
$\max \{ \sum_{i} (1+s_{i}t_{i})/2N, 1-\sum_{i} (1+s_{i}t_{i})/2N \}$, 
where $t_{i}$ is the planted group assignment such that $\{ t_{i} = +1 | \sigma_{i} = 1\}$ and $\{ t_{i} = -1 | \sigma_{i} = 2\}$. 
The minimum overlap is $0.5$ and is achieved when the group assignments are determined in a completely random manner. 
Owing to the finite size effect, the overlap only gradually decreases around the estimate of the detectability limit (solid yellow line). 
However, as shown in Fig.\ref{GreedyPhaseDiagram}(b), when the average degree is sufficiently high, the overlap decreases more sharply as $N$ increases, which implies that our estimate is accurate in the limit of $N \to \infty$. 
In the case of low average degrees, our result appears overestimated, likely because of the adopted approximations.

The notion of a metastable state is algorithm-dependent because it is defined with respect to a single spin flip. 
However, it is doubtful whether other update rules, such as cluster updates (i.e., multi-spin flips), may significantly improve performance in the case where the single spin flip algorithm (simple greedy algorithm) has a highly ragged modularity landscape.  
Therefore, it is worth comparing our estimated detectability limit with more sophisticated greedy algorithms. 


\textit{Detectability limit of Louvain algorithm}--- 
The Louvain algorithm \cite{Blondel2008} is a widely-used greedy heuristic for modularity maximization (see \cite{Blondel2008} for details regarding this algorithm). 
For the specific implementation, we used the code distributed at \cite{TraagLouvain}. 
The Louvain algorithm does not exactly correspond to the situation that we considered in the metastable state counting. 
First, the number of groups is determined automatically during the optimization process. 
Second, the Louvain algorithm contains multi-spin updates or cluster updates.

\begin{figure}[t!]
 \begin{center}
   \includegraphics[width= \columnwidth]{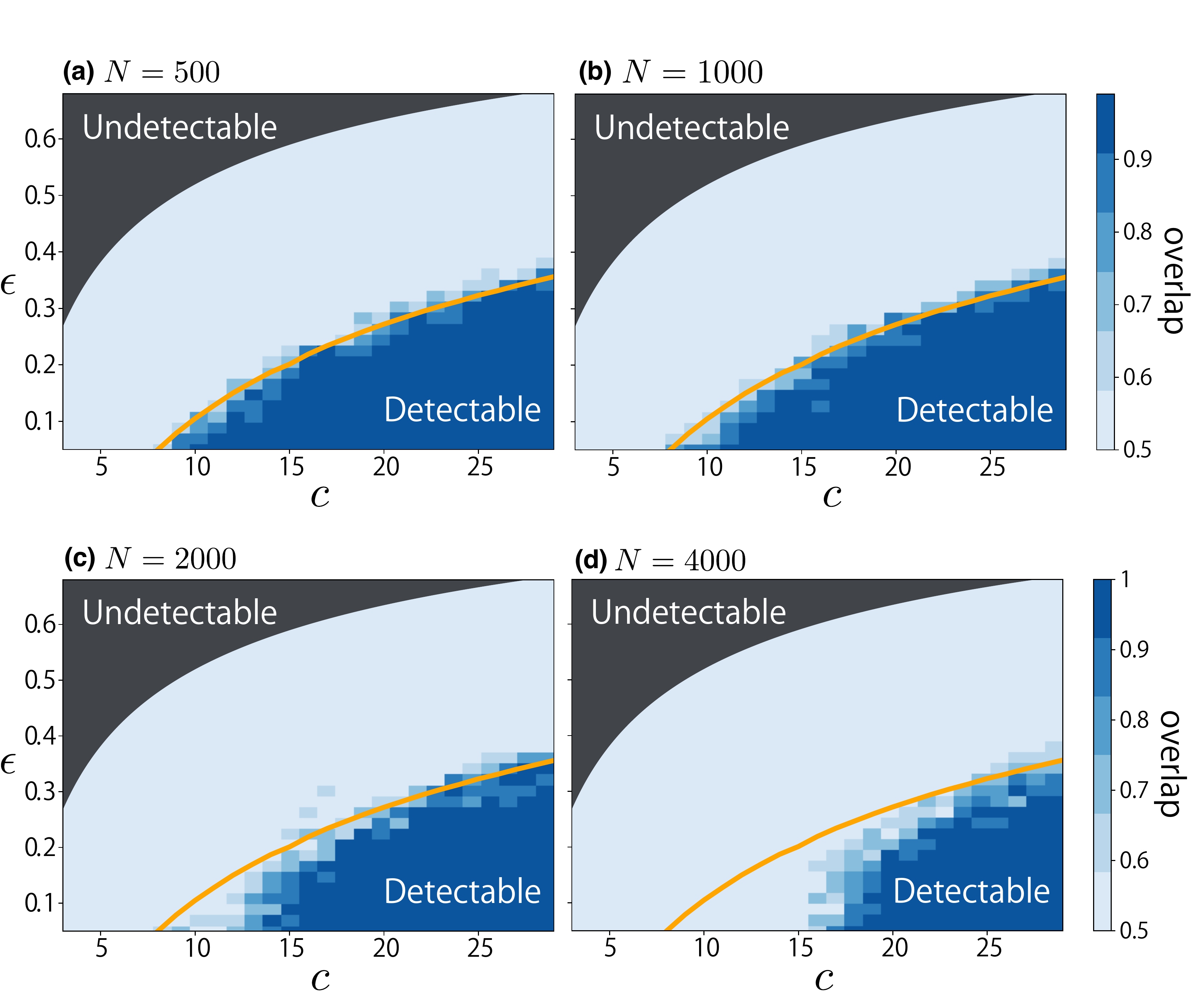}
 \end{center}
 \caption{
	(Color online) 
	Detectability phase diagrams of Louvain algorithm for a symmetric stochastic block model with (a) $N=500$, (b) $N=1000$, (c) $N=2000$, and (d) $N=4000$, respectively. 
	Plotting was carried out in the same manner as that shown in Fig.\ref{GreedyPhaseDiagram}. 
	The overlap is set to $0.5$ whenever the graph is partitioned into more than two groups. 
	In all plots, the average overlap value of ten graph instances was determined for each $c$ and $\epsilon$ pair.  
	}
 \label{LouvainPhaseDiagram}
\end{figure}

The detectability phase diagrams of the Louvain algorithm are shown in Fig.\ref{LouvainPhaseDiagram} as density plots. 
When the algorithm identifies more than two groups, we set the overlap to $0.5$. 

Interestingly, when the graph size $N$ is not very large, the detectability limit estimated by Eq.(\ref{TheDetectabilityLimit}) (solid yellow line) coincides with the phase boundary of the region where the overlap is greater than $0.5$, although the detectable region of the lower average degrees decreases as $N$ increases. 
To the extent of our investigation, the detectable region did not exceed Eq.(\ref{TheDetectabilityLimit}). 
This experimental observation implies that our estimate of the detectability limit is an intrinsic upper bound of modularity maximization, which holds more generally for greedy algorithms than for the single spin flip algorithm. 
The same analysis was carried out for the so-called \textit{fast greedy algorithm}, as presented in the Supplemental Material \cite{SupplementalMaterial}.

\textit{Discussion}--- 
Greedy algorithms have simple mechanisms and are relatively easy to implement. 
However, it is known that modularity maximization using a greedy algorithm is not optimal for inferring the stochastic block model. 
Here, we conducted a quantitative investigation with regard to this algorithm’s feasibility and limitations. 
Our result indicates that the algorithm fails for a considerably large region in the parameter space of the stochastic block model, even when the corresponding graphs have statistically significant structures. 
Note that we never focused on the true maximum of modularity; whether the partition with the maximum modularity is correlated to the planted partition is a very different problem and is not of our interest at all. 

Most importantly, our result indicates that greedy algorithms are expected to fail when a graph has a sufficiently low average degree, \textit{regardless of the modular structure’s strength}.
In the case of the symmetric stochastic block model, our approximated estimation predicted that this happens when $c \lesssim 7$. 
Although this value is not very accurate, our analysis successfully explains the experimentally observed limitations of the greedy algorithms in a qualitative manner. 
Thus, a quantitative insight into the limited utility of greedy algorithms is provided in terms of sparsity level. 
We also note that this limitation will be relaxed for the stochastic block models with different group sizes. This is because the symmetric stochastic block model, in which the group size is uninformative to the inference, is a relatively difficult problem.


When our objective is to extract meaningful structures from real-world networks, we should carefully investigate the behaviors of the considered algorithm. 
For example, while modularity maximization entails the risk of underfitting \cite{Fortunato2007}, it tends to overfit \cite{modBIX,Ghasemian2018} in comparison with other model selection criteria for various real-world networks. 
However, without quantitative knowledge, one might falsely expect a greedy algorithm to work well in a certain case, although there is very little chance that it will work appropriately. 

Note that the overfit and underfit concepts depend on the assumed graph ensemble, and many modern algorithms are formulated on the basis of the graph ensemble defined by the stochastic block model \cite{latouche12,ZhangMoore2014,PeixotoPRX2014,NewmanClauset2016,NewmanReinert2016,Peixoto2017tutorial}. 
Therefore, the present result can be used as a practical reference to perform modularity maximization. 


\textit{Acknowledgments}--- This study was funded by the New Energy and Industrial Technology Development Organization (NEDO) and JSPS KAKENHI (No. 18K18127 (TK) and No. 17H00764 (YK)).

\bibliographystyle{apsrev}
\bibliography{bib-greedy}

\clearpage

\setcounter{equation}{0}
\setcounter{figure}{0}
\setcounter{table}{0}
\setcounter{page}{1}
\makeatletter
\renewcommand{\theequation}{S\arabic{equation}}
\renewcommand{\thefigure}{S\arabic{figure}}
\renewcommand{\bibnumfmt}[1]{[S#1]}
\renewcommand{\citenumfont}[1]{S#1}

\begin{widetext}

{\bf Supplemental Material}

\section{On the definition of ``community structure''}
Because the algorithm aims to maximize modularity, one might think that the partition that achieves the true maximum of modularity should be the ``real'' community structure. 
In this study, we define the community structure as the planted group assignments of the stochastic block model. 
We only regard modularity maximization as an idea behind the algorithms. 

It is not appropriate to define the community structure as the true maximum of modularity because of the following reason. 
Although modularity maximization is equivalent to the inference using a particular type of stochastic block model, it should be emphasized that it is equivalent to the maximum likelihood estimate of the stochastic block model. 
Even for a graph generated by a uniformly random graph model, there always exists an optimum partition (which might be degenerated) of the graph in the sense of the maximum likelihood estimate. 
The maximum likelihood estimate in such a situation is nothing but an overfit and has no physical meaning. 
Moreover, because the idea of modularity function was to distinguish the actual graph from the instances of a uniform random graph model, the maximum likelihood estimate is undesirable also in the sense of modularity function.

\section{Derivation of the number of metastable states}
In this section, we present the detailed derivation of the number of metastable states $\mathcal{N}_{m}$. 
Using the step function $\Theta(x)$, where $\Theta(x>0)=1$ and $\Theta(x<0) = 0$, the number of metastable states can be counted as follows. 
\begin{align}
\mathcal{N}_{m} &= \sum_{\{s_{i}\}} \prod_{i} \int_{-\infty}^{\infty} d\lambda_{i} \Theta(\lambda_{i}) \, \delta\left( \lambda_{i} - s_{i} \sum_{j (\ne i)} B_{ij} s_{j} \right). \label{S-Nmetastable1}
\end{align}
The integral with respect to $\lambda_{i}$ from $0$ to $\infty$ can be recast using the integral representation of the step function. 
The delta functions can also be recast by the Fourier representation. 
Thus, we obtain the following expression.
\begin{align}
\mathcal{N}_{m} &= \sum_{\{s_{i}\}} \prod_{i} \int^{\infty}_{-\infty} \frac{d\hat{\lambda}_{i} d\lambda_{i}}{2\pi i} \frac{\mathrm{e}^{i \hat{\lambda_{i}}\lambda_{i}}}{\hat{\lambda}_{i} - i\epsilon} \, \int_{-\infty}^{\infty} \frac{d\phi_{i}}{2\pi} \, \mathrm{e}^{-i\phi_{i} (\lambda_{i} - s_{i} \sum_{j (\ne i)} B_{ij} s_{j})} \notag\\
&= \sum_{\{s_{i}\}} \int^{\infty}_{-\infty} \left(\prod_{i} \frac{d\phi_{i}}{2\pi i} \frac{1}{\phi_{i} - i\epsilon}\right) \, \mathrm{e}^{i\sum_{ij} \phi_{i} s_{i} B_{ij} s_{j}}. \label{S-Nmetastable2}
\end{align}
Specifically, the case where $i = j$ must be excluded for the sum in the exponent. 
However, we do not exclude it, because it only gives a vanishing contribution in the limit of $N \to \infty$. 

We denote $\left[ \cdots \right]_{A}$ for the quantities averaged over the graph ensemble of the stochastic block model. 
Here, we introduce the following order parameters.
\begin{align}
& F_{c} \equiv \frac{1}{N} \sum_{i=1}^{N} c_{i} s_{i}, \\
& G_{c} \equiv \frac{1}{N} \sum_{i=1}^{N} c_{i} s_{i} \phi_{i}, 
\end{align}
The ensemble average of $\mathcal{N}_{m}$ becomes 
\begin{align}
\left[ \mathcal{N}_{m} \right]_{A} 
&= \sum_{\{s_{i}\}} \int^{\infty}_{-\infty} \left(\prod_{i} \frac{d\phi_{i}}{2\pi i} \frac{1}{\phi_{i} - i\epsilon}\right) 
\mathrm{e}^{-i\alpha N^{2} F_{c} G_{c} } \notag\\
&\hspace{90pt}\times 
\left[ \mathrm{e}^{i \sum_{ij} A_{ij} \phi_{i}s_{i} s_{j}} \right]_{A}. \label{S-Nmetastable3}
\end{align}
Because the degree $c_{i}$ is constrained by $\ket{A}$ as $c_{i} = \sum_{j} A_{ij}$, it may seem that this condition should be treated within the ensemble average. 
However, the degree sequence only appears within the averaged quantities, namely, $F_{c}$ and $G_{c}$. 
Therefore, only the degree distribution is significant to the result, and asymptotically common for all graph instances. 
Therefore, we considered that all graph instances have the same degree sequence.

Subsequently, we calculate the ensemble average in Eq.(\ref{S-Nmetastable3}). 
The edges are generated independently and randomly in the stochastic block model, as follows.
\begin{align}
P(\{ A_{ij} \}) = \prod_{i<j} \rho_{\sigma_{i} \sigma_{j}}^{A_{ij}} \left(1 - \rho_{\sigma_{i} \sigma_{j}}\right)^{1-A_{ij}}. 
\end{align}
Thus, 
\begin{align}
\left[ \mathrm{e}^{i\sum_{ij} A_{ij} \phi_{i}s_{i}s_{j}} \right]_{A} 
&= \sum_{\{A_{ij}\}} P(\{ A_{ij} \}) \mathrm{e}^{i\sum_{ij} A_{ij} \phi_{i}s_{i}s_{j}} \notag\\ 
&= \prod_{i<j} \left( 1 + \rho_{\sigma_{i}\sigma_{j}} \left( \mathrm{e}^{i s_{i}s_{j}(\phi_{i} + \phi_{j})} - 1\right) \right) \notag\\
&\approx \exp \left[ \sum_{i<j} \rho_{\sigma_{i}\sigma_{j}} \left( \mathrm{e}^{i s_{i}s_{j}(\phi_{i} + \phi_{j})} - 1\right) \right]. \label{S-BoltzmannFactor1}
\end{align}
In the last line, we use the fact that $\rho_{\sigma \sigma^{\prime}} = O(N^{-1})$ for any $\sigma$ and $\sigma^{\prime}$. 

Moreover, we approximate that the magnitude obtained by $\phi_{i}$ is small. 
Recall that $\phi_{i}$ is the Fourier mode derived from the delta function with respect to $\lambda_{i}$ in Eq.(\ref{S-Nmetastable1}), and its integral ranges from negative infinity to positive infinity. 
Thus, neglecting the contribution of high frequency modes in the integral can be either interpreted as a low frequency approximation of the delta function or as the \textit{rotating wave approximation} \cite{FujiiRWA2017}, i.e., the contribution from the high frequency modes is approximately canceled out in the integral. This approximation is often adopted in the field of quantum mechanics. 
When regarded as a low frequency approximation, the present approximation is expected to be valid as long as the graphs are not extremely sparse and the modular structures are not very weak, because the point where $\lambda_{i}$ becomes nonzero in Eq.(\ref{S-Nmetastable1}) is typically far from zero. 
If the present approximation is valid in the sense of the rotating wave approximation, our estimation will be accurate even under a very sparse regime. 
Hence, we expand the exponent in Eq.(\ref{S-BoltzmannFactor1}) up to the second order in $\phi_{i}$ and $\phi_{j}$, i.e., 
\begin{align}
&\left[ \mathrm{e}^{i\sum_{ij} A_{ij} \phi_{i}s_{i}s_{j}} \right]_{A} \notag\\
&\hspace{10pt}\approx \exp \left[ \sum_{i<j} \rho_{\sigma_{i}\sigma_{j}} \left( -\frac{1}{2}(\phi_{i} + \phi_{j})^{2} + i s_{i}s_{j}(\phi_{i} + \phi_{j}) \right) \right]. \label{S-BoltzmannFactor2}
\end{align}

To break the coupling terms in Eq. (\ref{S-BoltzmannFactor2}), we introduce the following group-wise order parameters. 
\begin{align}
& E_{\sigma} \equiv \frac{1}{\gamma_{\sigma}N} \sum_{i \in V_{\sigma}} \phi_{i}, \\
& F_{\sigma} \equiv \frac{1}{\gamma_{\sigma}N} \sum_{i \in V_{\sigma}} s_{i}, \\
& G_{\sigma} \equiv \frac{1}{\gamma_{\sigma}N} \sum_{i \in V_{\sigma}} s_{i} \phi_{i}, \\
& H_{\sigma} \equiv \frac{1}{\gamma_{\sigma}N} \sum_{i \in V_{\sigma}} \phi^{2}_{i}.
\end{align}
Then,
Eq.(\ref{S-BoltzmannFactor2}) reads as follows. 
\begin{align}
& \exp \left[ \frac{1}{2} \sum_{\sigma \sigma^{\prime}} \rho_{\sigma \sigma^{\prime}} 
\sum_{i \in V_{\sigma}} \sum_{j \in V_{\sigma^{\prime}}} 
\left( -\frac{1}{2}(\phi^{2}_{i} + \phi^{2}_{j}) -\phi_{i}\phi_{j} + i s_{i}s_{j}(\phi_{i} + \phi_{j}) \right) \right] \\
&= \exp \Biggl[ \frac{N}{2} \sum_{\sigma \sigma^{\prime}} B_{\sigma \sigma^{\prime}} 
\left( -\frac{1}{2} \left( H_{\sigma} + H_{\sigma^{\prime}} \right) 
- E_{\sigma}E_{\sigma^{\prime}} 
+ i\left( F_{\sigma}G_{\sigma^{\prime}} + F_{\sigma^{\prime}}G_{\sigma} \right)
\right) \Biggr], 
\end{align}
and 
\begin{align}
\left[ \mathcal{N}_{m} \right]_{A} 
&= \sum_{\{s_{i}\}} \int \left(\prod_{i} \frac{d\phi_{i}}{2\pi i} \frac{1}{\phi_{i} - i\epsilon}\right) \int N\frac{d\hat{F}_{c}dF_{c}}{2\pi i} \int N \frac{d\hat{G}_{c}dG_{c}}{2\pi} 
\int \left( \prod_{\sigma} \gamma_{\sigma}N \frac{d\hat{E}_{\sigma}dE_{\sigma}}{2\pi} \, \frac{d\hat{F}_{\sigma}dF_{\sigma}}{2\pi i} \, \frac{d\hat{G}_{\sigma}dG_{\sigma}}{2\pi} \, \frac{d\hat{H}_{\sigma}dH_{\sigma}}{4\pi i} \right) \notag\\
&\hspace{10pt}\times \exp\Biggl[ 
-i\alpha N^{2} F_{c} G_{c} 
+ \frac{N}{2} \sum_{\sigma \sigma^{\prime}} B_{\sigma \sigma^{\prime}} 
\left( -\frac{1}{2} \left( H_{\sigma} + H_{\sigma^{\prime}} \right) 
- E_{\sigma}E_{\sigma^{\prime}} 
+ i\left( F_{\sigma}G_{\sigma^{\prime}} + F_{\sigma^{\prime}}G_{\sigma} \right)
\right) \notag\\
&\hspace{40pt}- \hat{F}_{c} \left( N F_{c} - \sum_{i} c_{i} s_{i} \right) 
- i\hat{G}_{c} \left( N G_{c} - \sum_{i} c_{i} s_{i} \phi_{i} \right) \notag\\
&\hspace{40pt}- i\sum_{\sigma} \hat{E}_{\sigma} \left( \gamma_{\sigma}N E_{\sigma} - \sum_{i \in V_{\sigma}} \phi_{i} \right) 
- \sum_{\sigma} \hat{F}_{\sigma} \left( \gamma_{\sigma}N F_{\sigma} - \sum_{i \in V_{\sigma}} s_{i} \right) \notag\\
&\hspace{40pt}- i\sum_{\sigma} \hat{G}_{\sigma} \left( \gamma_{\sigma}N G_{\sigma} - \sum_{i \in V_{\sigma}} s_{i}\phi_{i} \right) 
+ \frac{1}{2} \sum_{\sigma} \hat{H}_{\sigma} \left( \gamma_{\sigma}N H_{\sigma} - \sum_{i \in V_{\sigma}} \phi^{2}_{i} \right) 
\Biggr]. \label{S-Nmetastable4}
\end{align}

The integral with respect to $\phi_{i}$ in Eq. (\ref{S-Nmetastable4}) can be performed in a straightforward manner, as follows. 
For $i \in V_{\sigma}$, 
\begin{align}
& \int^{\infty}_{-\infty} \frac{d\phi_{i}}{2\pi i} \frac{1}{\phi_{i} - i\epsilon} \mathrm{e}^{-\frac{1}{2}\hat{H}_{\sigma} \phi^{2}_{i} + i(\hat{E}_{\sigma} + s_{i}\hat{G}_{\sigma} + c_{i}s_{i}\hat{G}_{c})\phi_{i} } \notag\\
&= \int^{\hat{H}^{-1/2}_{\sigma}(\hat{E}_{\sigma} + s_{i}\hat{G}_{\sigma} + c_{i}s_{i}\hat{G}_{c})}_{-\infty} \frac{dx}{\sqrt{2\pi}} \mathrm{e}^{-\frac{1}{2} x^{2}} \notag\\
&= \Phi\left( \hat{H}^{-\frac{1}{2}}_{\sigma}(\hat{E}_{\sigma} + s_{i}\hat{G}_{\sigma} + c_{i}s_{i}\hat{G}_{c}) \right), 
\end{align}
where $\Phi(\cdot)$ is the standard normal cumulative distribution function defined in the main text. 

We denote the ``partition function'' with respect to the spin variable $s_{i}$, as follows.
\begin{align}
Z_{c_{i}\sigma} &\equiv 
\sum_{s_{i}} \mathrm{e}^{s_{i}( \hat{F}_{\sigma} + c_{i}\hat{F}_{c} )} 
\Phi\left( \hat{H}^{-\frac{1}{2}}_{\sigma}(\hat{E}_{\sigma} + s_{i}\hat{G}_{\sigma} + c_{i}s_{i}\hat{G}_{c}) \right), 
\end{align}
Additionally, we let the degree distribution of the vertices in the group be $\mathcal{P}_{\sigma}(k)$, which has the form of the Poisson distribution in the infinite graph limit. 
Then, 
\begin{align}
\prod_{i} Z_{c_{i}\sigma_{i}} 
&= \exp \left( \sum_{\sigma} \sum_{i \in V_{\sigma}} \log Z_{c_{i}\sigma} \right) \notag\\
&= \exp \left( N \sum_{\sigma} \gamma_{\sigma} \sum_{k} \mathcal{P}(k) \log Z_{k\sigma} \right), 
\end{align}
and $\left[ \mathcal{N}_{m} \right]_{A}$ becomes
\begin{align}
\left[ \mathcal{N}_{m} \right]_{A} 
&= \int N\frac{d\hat{F}_{c}dF_{c}}{2\pi i} \int N \frac{d\hat{G}_{c}dG_{c}}{2\pi} 
\int \prod_{\sigma} \gamma_{\sigma}N \left( \frac{d\hat{E}_{\sigma}dE_{\sigma}}{2\pi} \, \frac{d\hat{F}_{\sigma}dF_{\sigma}}{2\pi i} \, \frac{d\hat{G}_{\sigma}dG_{\sigma}}{2\pi} \, \frac{d\hat{H}_{\sigma}dH_{\sigma}}{4 \pi i} \right) \, \mathrm{e}^{N f}, \\
-f &= 
i\alpha N F_{c} G_{c} 
+\frac{1}{2} \sum_{\sigma \sigma^{\prime}} B_{\sigma \sigma^{\prime}} 
\left[ \frac{1}{2} \left( H_{\sigma} + H_{\sigma^{\prime}} \right) 
+ E_{\sigma}E_{\sigma^{\prime}} 
-i\left( F_{\sigma}G_{\sigma^{\prime}} + F_{\sigma^{\prime}}G_{\sigma} \right)
\right] \notag\\
&\hspace{10pt}+\hat{F}_{c} F_{c} + i\hat{G}_{c} G_{c} + \sum_{\sigma} \gamma_{\sigma} \left( i\hat{E}_{\sigma} E_{\sigma} + \hat{F}_{\sigma} F_{\sigma} + i\hat{G}_{\sigma} G_{\sigma} - \frac{1}{2}\hat{H}_{\sigma} H_{\sigma} - \sum_{k} \mathcal{P}(k) \log Z_{k\sigma} \right). \label{S-Nmetastable5}
\end{align}

\subsection{Saddle-point conditions}
In the limit of $N \to \infty$, the integral of Eq.(\ref{S-Nmetastable5}) can be evaluated by its saddle-point estimate. 
From the saddle-point conditions of $f$, we have $\hat{H}_{\sigma} = c_{\sigma}$, where $c_{\sigma}$ is the average degree of a vertex belonging to $V_{\sigma}$ and the following self-consistent equations. 
\begin{align}
\hat{F}_{c} &= -\alpha N \sum_{\sigma} \frac{\gamma_{\sigma}}{\sqrt{c_{\sigma}}} \sum_{k} \mathcal{P}_{\sigma}(k) k \bracket{s \Psi_{k\sigma}(s)}_{Z_{k\sigma}} \label{S-selfconsistent-1}\\
\hat{G}_{c} &= -\alpha N \sum_{\sigma} \gamma_{\sigma} \sum_{k} \mathcal{P}_{\sigma}(k) k \bracket{s}_{Z_{k\sigma}},  \label{S-selfconsistent-2}\\
\hat{E}_{\sigma} &= \sum_{\sigma^{\prime}} \frac{B_{\sigma \sigma^{\prime}}}{\gamma_{\sigma} \sqrt{c_{\sigma^{\prime}}}} \sum_{k} \mathcal{P}_{\sigma^{\prime}}(k) \, \bracket{\Psi_{k\sigma^{\prime}}(s)}_{Z_{k\sigma^{\prime}}},  \label{S-selfconsistent-3}\\
\hat{F}_{\sigma} &= \sum_{\sigma^{\prime}} \frac{B_{\sigma \sigma^{\prime}}}{\gamma_{\sigma} \sqrt{c_{\sigma^{\prime}}}} \sum_{k} \mathcal{P}_{\sigma^{\prime}}(k) \bracket{s \Psi_{k\sigma^{\prime}}(s)}_{Z_{k\sigma^{\prime}}}  \label{S-selfconsistent-4}\\
\hat{G}_{\sigma} &= \sum_{\sigma^{\prime}} \frac{B_{\sigma \sigma^{\prime}}}{\gamma_{\sigma}} \sum_{k} \mathcal{P}_{\sigma^{\prime}}(k) \, \bracket{s}_{Z_{k\sigma^{\prime}}},  \label{S-selfconsistent-5}
\end{align}
where 
\begin{align}
& \Psi_{k\sigma}(s) = \frac{1}{\sqrt{2\pi}} \frac{\mathrm{e}^{-\frac{1}{2c_{\sigma}} (\hat{E}_{\sigma} + s\hat{G}_{\sigma} + ks\hat{G}_{c})^{2}}}{\Phi\left( \frac{\hat{E}_{\sigma} + s\hat{G}_{\sigma} + ks\hat{G}_{c}}{\sqrt{c_{\sigma}}} \right)}, \\
& \bracket{X(s)}_{Z_{k\sigma}} \equiv \frac{1}{Z_{k\sigma}} \sum_{s} X(s) \mathrm{e}^{s( \hat{F}_{\sigma} + k\hat{F}_{c} )} \notag\\ 
&\hspace{60pt}\times \Phi\left( \frac{\hat{E}_{\sigma} + s\hat{G}_{\sigma} + ks\hat{G}_{c}}{\sqrt{c_{\sigma}}} \right). 
\end{align}
The function $f$ is expressed in terms of these variables as follows.
\begin{align}
f &= \sum_{\sigma} \gamma_{\sigma} \left( \frac{i}{2}\hat{E}_{\sigma} E_{\sigma} - c_{\sigma} H_{\sigma} + \sum_{k} \mathcal{P}_{\sigma}(k) \log Z_{k\sigma} \right), \label{S-FreeEnergy}
\end{align}
where 
\begin{align}
i E_{\sigma} &= \frac{1}{\sqrt{c_{\sigma}}} \sum_{k} \mathcal{P}_{\sigma}(k) \, \bracket{\Psi_{k\sigma}(s)}_{Z_{k\sigma}}, \\
H_{\sigma} &= c^{-3/2}_{\sigma} \hat{E}_{\sigma} \sum_{k} \mathcal{P}_{\sigma}(k) \, \bracket{\Psi_{k\sigma}(s)}_{Z_{k\sigma}} \notag\\
&+ \sum_{k} \mathcal{P}_{\sigma}(k) \, (\hat{G}_{\sigma} + k \hat{G}_{c}) \bracket{s \Psi_{k\sigma}(s)}_{Z_{k\sigma}}.
\end{align}


\subsection{Symmetric stochastic block model}
The above equations refer to a general two-group stochastic block model. 
Let us consider a more specific case by approximating further to obtain a set of equations with a more compact form. 
We assume that the symmetric stochastic block model, i.e., $\gamma_{\sigma} = 1/2$ and $c_{\sigma} = c$ for both $\sigma$, and the matrix $\ket{B} = [B_{\sigma \sigma^{\prime}}]$ are parametrized as follows.
\begin{align}
\ket{B} 
= \frac{c}{2(1+\epsilon)} 
\begin{pmatrix}
1 & \epsilon \\
\epsilon & 1
\end{pmatrix}. 
\end{align}
In this case, the order parameters are either symmetric or identical for the different values of $\sigma$. 
Moreover, we set the resolution parameter to a commonly used value, $\alpha = 1/cN$, and adopt the regular approximation, i.e., $\mathcal{P}(k) \approx \delta_{k,c}$. 

We define 
\begin{align}
&\tilde{F}_{\sigma} \equiv \hat{F}_{\sigma} + c\hat{F}_{c}, \\
&\tilde{G}_{\sigma} \equiv \hat{G}_{\sigma} + c\hat{G}_{c}.
\end{align}
From the symmetry condition, we have $\tilde{F}_{1} = -\tilde{F}_{2}$ and $\tilde{G}_{1} = -\tilde{G}_{2}$. 
From the saddle-point condition, it can be seen that $\hat{E}_{\sigma}$ is positive for any $\sigma$, i.e., $\hat{E}_{1} = \hat{E}_{2}$. 
Then, the set of self-consistent equations, namely, the set containing Eqs.(\ref{S-selfconsistent-1})--(\ref{S-selfconsistent-5}) and Eq.(\ref{S-FreeEnergy}), is simplified as described in the main text.

\section{Further numerical confirmation on the performance of the simple greedy algorithm and Louvain algorithm}\label{Appendix-LouvainDetail}
In Fig.~2(b) in the main text, we presented the mean overlaps around the detectability limit. 
Here, in Fig.~\ref{LouvainDetectability}(a), we instead present the median overlaps and interquartile ranges of ten graph instances. 
Although the mean overlaps remained high near the estimated detectability limit, in fact, we can confirm that the fluctuation of the overlap values becomes considerable near the estimated limit. 

The same analysis is done for the overlaps obtained by the Louvain algorithm, corresponding to Fig.~3 in the main text. 
Here, in Fig.~\ref{LouvainDetectability}(b), we present the medians overlaps and interquartile ranges of ten graph instances with respect to a few specified average degrees. 
We can confirm that our estimate of the detectability limit is indeed very accurate when the average degree is relatively large. 

\begin{figure}[h!]
 \begin{center}
   \includegraphics[width= 0.99\columnwidth]{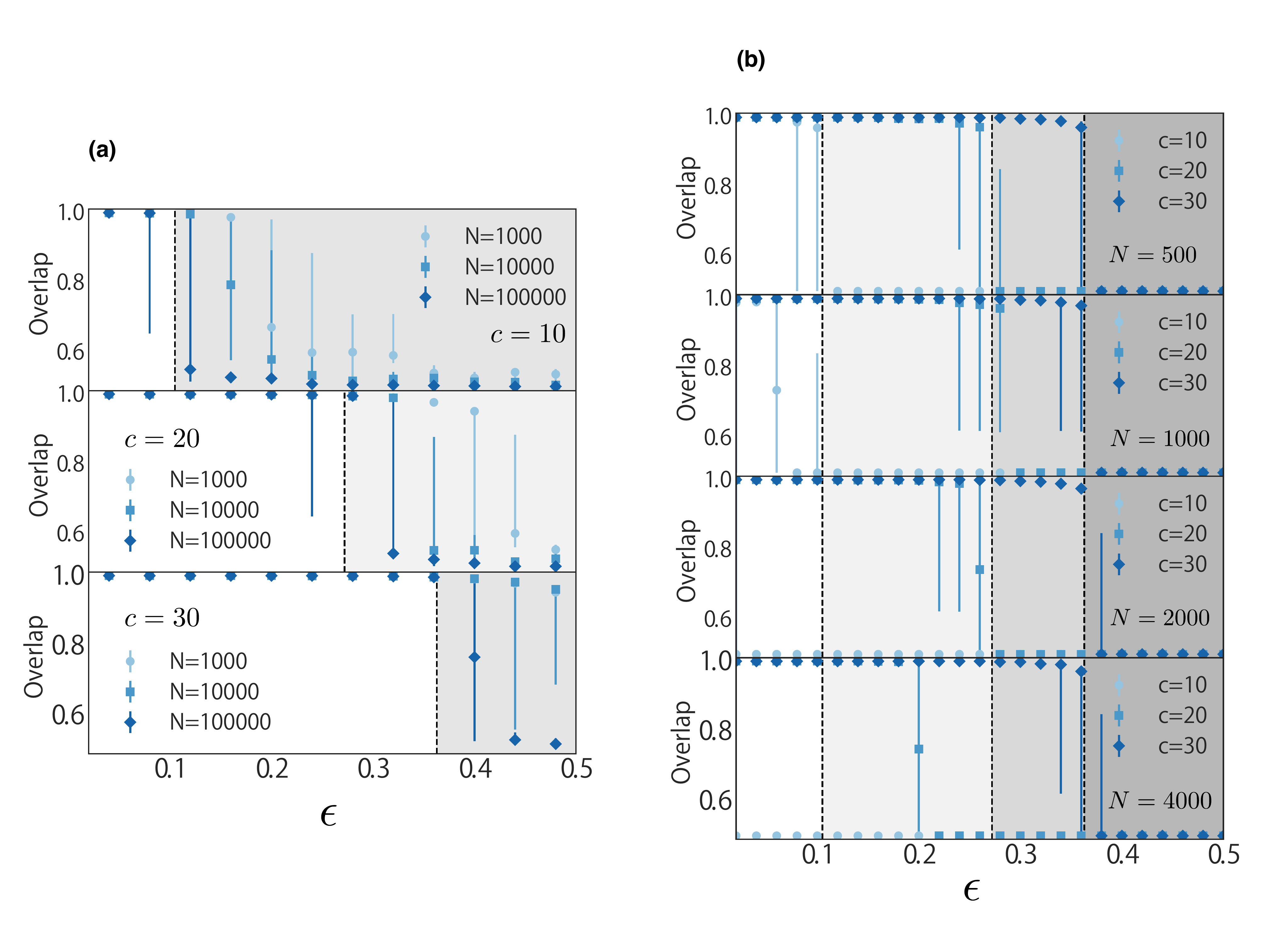}
 \end{center}
 \caption{
	(Color online) 
	(a) Overlaps obtained by the simple greedy algorithm. 
	From top to bottom, the results with $c=10$, $c=20$, and $c=30$ are presented. 
	Each panel represents the results of different graph sizes $N$ as functions of $\epsilon$. 
	The shaded regions with the dashed border lines represent our undetectable region estimates. 
	(b) Overlaps obtained by the Louvain algorithm. 
	From top to bottom, the results with different graph sizes $N$ are presented. 
	Each panel represents the results of $c = 10$, $c = 20$, and $c = 30$ as functions of $\epsilon$. 
	The shaded regions with the dashed border lines represent, from the left to right, our undetectable region estimates for $c = 10$, $20$, and $30$, respectively. 
	In all panels, the points represent the median overlaps and the error bars represent the interquartiles with respect to the results of ten graph instances. 
	}
 \label{LouvainDetectability}
\end{figure}

\section{Detectability phase diagrams of fast greedy algorithm}\label{Appendix-FastGreedy}

\begin{figure}[h!]
 \begin{center}
   \includegraphics[width= \columnwidth]{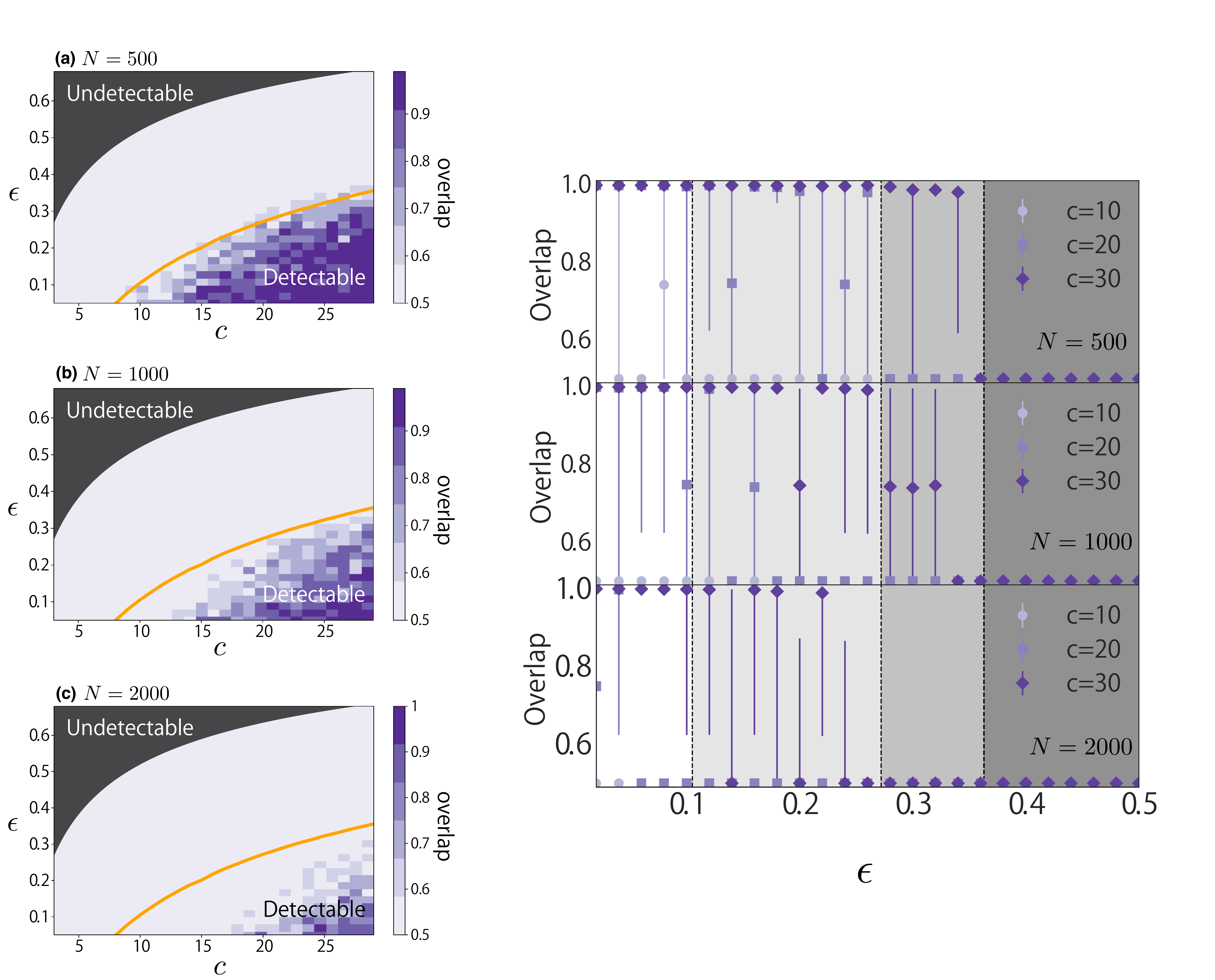}
 \end{center}
 \caption{
	(Color online) 
	(Left)
	Detectability phase diagrams of fast greedy algorithm for symmetric stochastic block model with different graph sizes $N$. 
	The diagrams are plotted in the same manner as those shown in Fig.~3 in the main text. 
	Graphs with (a) $N=500$, (b) $1000$, and (c) $2000$ are presented. 
	In all plots, the average overlap value of ten graph instances is determined for each pair of $c$ and $\epsilon$. 
	(Right) 
	Overlaps of $c = 10$, $c = 20$, and $c = 30$ as functions of $\epsilon$ for different graph sizes $N$. 
	They are plotted in the same manner as those shown in Fig.~\ref{LouvainDetectability}. 
	}
 \label{fastgreedyPhaseDiagrams}
\end{figure}

In this section, we consider the performance of the \textit{fast greedy algorithm} \cite{Clauset2004} (see \cite{Clauset2004} for a detailed description of the algorithm). 
This is a simpler implementation of modularity maximization, in comparison with the Louvain algorithm.
For this specific implementation, we use the code embedded in \textit{python-igraph} \cite{pythonigraphFastgreedy}. 
As in the case of the Louvain algorithm, the number of groups is not given as input, but is automatically estimated during the optimization process. 

The detectability phase diagrams of the fast greedy algorithm for the symmetric stochastic block model are presented in Fig.\ref{fastgreedyPhaseDiagrams}. 
Again, the overlap was set to $0.5$ whenever the graph was partitioned into more than two groups. 
It can be observed that, for $N=1000$, the fast greedy algorithm does not achieve the estimated limit, while the Louvain algorithm (almost) achieves it (see Fig.~3(b) in the main text). 
Conversely, for the graph instances with a smaller size ($N=500$) shown in Fig. \ref{fastgreedyPhaseDiagrams}(a), the fast greedy algorithm works up to the estimated detectability limit. 
As in the case of the Louvain algorithm, it can be seen in Figs.~\ref{fastgreedyPhaseDiagrams}b and c that sparser graphs are significantly affected by the considered graph size. 

\end{widetext}

\end{document}